\documentclass[aps,pra,reprint,superscriptaddress,showkeys,footinbib,floatfix]{revtex4-1}

\usepackage[T1]{fontenc}			
\usepackage[utf8]{inputenc}			
\usepackage[english]{babel}			
\usepackage{microtype}				

\usepackage{color} 						
\definecolor{red}{rgb}{1,0,0}					
\definecolor{blue}{rgb}{0,0,1}					
\definecolor{black}{rgb}{0,0,0}				

\usepackage{soul} 
\definecolor{hlyellow}{rgb}{0.95,0.95,0}
\definecolor{hlgreen}{rgb}{0,0.95,0}
\sethlcolor{hlyellow}
\soulregister \ref 7
\soulregister \cite 7
\soulregister \citet 7
\soulregister \footnote 8
\soulregister {\ } 7
\soulregister \figref 7
\soulregister \tabref 7
\soulregister \secref 7
\soulregister \refref 7
\soulregister \eqref 7

\usepackage{amsmath}	
\usepackage{amssymb,amsfonts,mathrsfs} 
\usepackage{array} 

\usepackage{graphicx}			

\usepackage{hyperref}  
\definecolor{dullmagenta}{rgb}{0.4,0,0.4}   
\definecolor{darkblue}{rgb}{0,0,0.4}
\definecolor{medblue}{rgb}{0,0,0.6}
\definecolor{lightblue}{rgb}{0,0,0.8}
\hypersetup{
	colorlinks=true, 		
	breaklinks=true,		
	linkcolor=lightblue,
	citecolor=lightblue,
	urlcolor=lightblue,
	filecolor=dullmagenta
}
\usepackage[all]{hypcap} 	

\usepackage{fixgreek}
\usepackage{physmaths}
\usepackage{physunits}
\usepackage{multirow} 

\newcommand{\figpdfa}{-pdfa} 
\newcommand{\figref}[1]{Fig.\ \ref{#1}} 
\newcommand{\tabref}[1]{Tab.\ \ref{#1}} 
\newcommand{\secref}[1]{Sec.\ \ref{#1}} 
\newcommand{\refref}[1]{Ref.\ \cite{#1}} 
\newcommand{\eqnref}[1]{Eq.\ \eqref{#1}} 
\newcommand{\figletter}[1]{\textbf{(\mbox{#1})}}

\newcommand{\SiOtwo}{SiO$_2$}
\newcommand{\twentyeightSi}{$^{28}$Si}
\newcommand{\twentynineSi}{$^{29}$Si}
\newcommand{\ueVpT}{\text{ }\upmu\text{eV}\,\text{T}^{-1}}

\newcommand{\DEz}{\Delta E_z}
\newcommand{\bigket}[1]{\big\vert #1 \big\rangle}
\newcommand{\bigbra}[1]{\big\langle #1 \big\vert}

\newcommand{\supmat}{Supplemental Material}
\newcommand{\SupRefSecSapRap}{\secref{sec:saprap}} 
\newcommand{\SupRefSecPulseParams}{\secref{sec:pulseparams}} 
\newcommand{\SupRefSecSTMinus}{\secref{sec:stminussup}} 
\newcommand{\SupRefSecRelax}{\secref{sec:trelax}} 
\newcommand{\SupRefSecDNP}{\secref{sec:dnp}} 
\newcommand{\SupRefSecModel}{\secref{sec:model}} 
\newcommand{\SupEqnOmegaSOfit}{\eqnref{eq:omegasofit}} 

\begin{document}

\newcommand{\mytitle}
{Spin-orbit Interactions for Singlet-Triplet Qubits in Silicon}
\title{\mytitle}

\author{Patrick \surname{Harvey-Collard}}
\email[Correspondence to: ]{P.Collard@USherbrooke.ca}
\affiliation{Département de physique et Institut quantique, Université de Sherbrooke, Sherbrooke (Québec) J1K 2R1, Canada}
\affiliation{Sandia National Laboratories, Albuquerque, New Mexico 87185, USA}

\author{N.~Tobias \surname{Jacobson}}
\affiliation{Center for Computing Research, Sandia National Laboratories, Albuquerque, New Mexico 87185, USA}

\author{Chloé \surname{Bureau-Oxton}}
\affiliation{Département de physique et Institut quantique, Université de Sherbrooke, Sherbrooke (Québec) J1K 2R1, Canada}
\affiliation{Sandia National Laboratories, Albuquerque, New Mexico 87185, USA}

\author{Ryan~M. \surname{Jock}}
\affiliation{Sandia National Laboratories, Albuquerque, New Mexico 87185, USA}

\author{Vanita \surname{Srinivasa}}
\affiliation{Center for Computing Research, Sandia National Laboratories, Albuquerque, New Mexico 87185, USA}

\author{Andrew~M. \surname{Mounce}}
\affiliation{Sandia National Laboratories, Albuquerque, New Mexico 87185, USA}



\author{Daniel~R. \surname{Ward}}
\affiliation{Sandia National Laboratories, Albuquerque, New Mexico 87185, USA}
\author{John~M. \surname{Anderson}}
\affiliation{Sandia National Laboratories, Albuquerque, New Mexico 87185, USA}
\author{Ronald~P. \surname{Manginell}}
\affiliation{Sandia National Laboratories, Albuquerque, New Mexico 87185, USA}
\author{Joel~R. \surname{Wendt}}
\affiliation{Sandia National Laboratories, Albuquerque, New Mexico 87185, USA}
\author{Tammy \surname{Pluym}}
\affiliation{Sandia National Laboratories, Albuquerque, New Mexico 87185, USA}

\author{Michael~P. \surname{Lilly}}
\affiliation{Center for Integrated Nanotechnologies, Sandia National Laboratories, Albuquerque, New Mexico 87185, USA}

\author{Dwight~R. \surname{Luhman}}
\affiliation{Sandia National Laboratories, Albuquerque, New Mexico 87185, USA}

\author{Michel \surname{Pioro-Ladrière}}
\affiliation{Département de physique et Institut quantique, Université de Sherbrooke, Sherbrooke (Québec) J1K 2R1, Canada}
\affiliation{Quantum Information Science Program, Canadian Institute for Advanced Research, Toronto (Ontario) M5G 1Z8, Canada}

\author{Malcolm~S. \surname{Carroll}}
\email[Correspondence to: ]{mscarro@sandia.gov}
\affiliation{Sandia National Laboratories, Albuquerque, New Mexico 87185, USA}

\date{May 13, 2019}

\begin{abstract}
Spin-orbit coupling is relatively weak for electrons in bulk silicon, but enhanced interactions are reported in nanostructures such as the quantum dots used for spin qubits. These interactions have been attributed to various dissimilar interface effects, including disorder or broken crystal symmetries. In this Letter, we use a double-quantum-dot qubit to probe these interactions by comparing the spins of separated singlet-triplet electron pairs. We observe both intravalley and intervalley mechanisms, each dominant for [110] and [100] magnetic field orientations, respectively, that are consistent with a broken crystal symmetry model. We also observe a third spin-flip mechanism caused by tunneling between the quantum dots. This improved understanding is important for qubit uniformity, spin control and decoherence, and two-qubit gates.
\end{abstract}

\maketitle


\section{Introduction} 

Isotopically enriched silicon is a prime semiconductor for the implementation of spin qubits \cite{loss1998}. In addition to reduced spin decoherence enabled by the near absence of lattice nuclear spins \cite{witzel2010,veldhorst2014a}, silicon is a low spin-orbit coupling material for electrons that enables long spin relaxation times \cite{zwanenburg2013,watson2017} and low coupling to charge noise. In silicon quantum dots (QDs), recent work has shown that spin-orbit effects arise in the presence of strong electron confinement \cite{yang2013b,veldhorst2015b,eng2015,ferdous2018a,jock2018,corna2018}. This enhanced interaction has been attributed to intervalley spin-orbit coupling and interface disorder \cite{yang2013b,veldhorst2015b,corna2018} in some works, and to broken crystal symmetries \cite{rossler2002} at the Si/\SiOtwo{} \cite{jock2018} or Si/SiGe \cite{prada2011} interfaces in other works. Recently, \citet{jock2018} have used a singlet-triplet (ST) qubit \cite{levy2002,petta2005} to probe the electron $g$-factor difference between two QDs, and found a strong magnetic-field-dependent anisotropy explained with an intravalley mechanism. This anisotropy can be exploited to enhance spin-orbit effects for spin control \cite{veldhorst2014a,jock2018}, or suppress them for uniformity and reproducibility \cite{li2018a}. 
ST qubits are promising candidates for quantum computing, thanks to the ability to perform exchange \cite{levy2002}, capacitive \cite{shulman2012a,srinivasa2015a,nichol2017}, and long-range \cite{trifunovic2012,mehl2014,srinivasa2015,malinowski2018} two-qubit gates, as well as low-frequency one-qubit operations \cite{taylor2007}. In GaAs devices, the use of differential dynamic nuclear polarization (DDNP) was shown to dramatically enhance the ST qubit coherence \cite{bluhm2010,shulman2014}, and also enable its control \cite{foletti2009,shulman2014}. The DDNP technique depends on the interaction between the two-electron spin singlet $\ket{S} = (\ket{\uparrow\downarrow}-\ket{\downarrow\uparrow})/\sqrt{2}$ and spin triplet $\ket{T_-} = \ket{\downarrow\downarrow}$ mediated by the hyperfine coupling to lattice nuclear spins \cite{gullans2010}. It was shown that spin-orbit interaction can couple these two states as well \cite{stepanenko2012}, impacting the ability to perform a DDNP by providing an alternate channel to dissipate angular momentum \cite{rancic2014,nichol2015a}. 
In light of these different works, it remains unclear what spin-orbit effects predominate in different situations, what their microscopic origins are, and how these effects will impact the operation of silicon devices.

In this Letter, we report the observation of three different spin-orbit effects in the same device using a ST qubit in isotopically enriched silicon. The first two effects are probed using $S{-}T_0$ precession and appear at different orders of perturbation theory. They consist of an intravalley $g$-factor difference effect and an intervalley spin-coupling effect. The dominant mechanism depends on the magnetic field orientation with respect to crystallographic axes. We report here a nonlinear magnetic field strength dependence, in addition to previously reported linear dependences. The third effect is probed using $S{-}T_-$ spin transitions and involves a spin-flip process triggered by electron tunneling between the QDs. To measure this effect, we adapt a method previously used in GaAs \cite{nichol2015a} to our silicon system, where the near absence of nuclear spins otherwise prevents these transitions. We find that the enhanced spin-orbit interaction in the device strongly couples these states, as it does for GaAs devices. In fact, the spin-orbit length estimated from our measurements is only slightly smaller than bulk GaAs values, a result that is in accordance with other recent observations of strong spin-orbit effects in silicon nanodevices. This prevents us from performing DDNP of the residual \twentynineSi{}. 

The effects are modeled with an analytical microscopic intravalley theory based on broken crystal symmetries introduced in \citet{jock2018} and extended in this Letter to describe the additional intervalley effect reported here. The model involves the electron momentum only at the Si--\SiOtwo{} interface, resulting in stronger-than-bulk first-order effects in the electron momentum and clear predictions that could help elucidate the microscopic origin of the enhanced spin-orbit effects in the future \cite{ruskov2018}. This Letter, as a consequence of its comprehensive view of spin orbit interactions, will affect how pulses are shaped around the uncovered transitions in silicon qubits as well as providing more detailed guidance about the implications of how samples are mounted in dilution refrigerators with respect to magnetic fields.

\section{Methods}

\begin{figure}[tbp]
   \centering
   \includegraphics{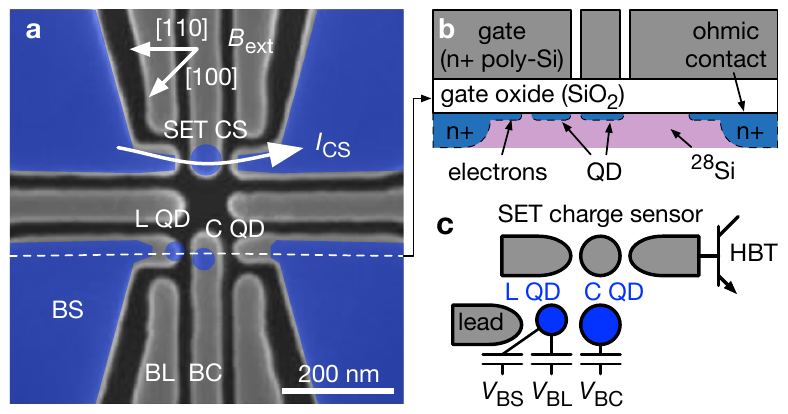} 
   \caption{\figletter{a} Scanning electron microscope image of the gate structure of the silicon QD device. The blue overlay indicates the estimated locations of electron accumulation. The crystallographic and external magnetic field ($B_\text{ext}$) orientations are indicated. All experiments are performed with the magnetic field along the [100] orientation, except when otherwise specified. A SET CS is used for readout and sensing. Its current $I_\text{CS}$ is amplified using a SiGe HBT.     \figletter{b} Schematic lateral view of the device structure (along the dashed line).     \figletter{c} Conceptual view of the two QDs, the single-lead reservoir, the CS and the predominant QD-gate capacitive couplings. The BL and BC gates are used for fast electrical control of the L and C QDs, respectively.}
   \label{fig:device}
\end{figure}

The experiments are performed in a dilution refrigerator with an electronic temperature of around $300 \mK$. The gated silicon QD device is shown in \figref{fig:device}. The silicon is isotopically enriched \twentyeightSi{}, with a measured $685 \ppm$ of residual \twentynineSi{}. Fabrication and device crystallographic orientation are as in \citet{jock2018}; the device is from the same fabrication run but a different die and measured in a different system. Two QDs are formed, one under the bottom source (BS) gate and one under the bottom center (BC) gate. The bottom left (BL) and BC gates are used for fast control of the left (L) and center (C) QD charge occupations $(N_\text{L}, N_\text{C})$ and interdot detuning $\eps$.

The double QD is biased in a two-electron charge configuration to form a ST qubit. The L QD has a ST splitting of $243 \ueV$ and the C QD has one of $185 \ueV$ (the latter measured in a $(3,2)$ configuration to avoid charge latching). Spin readout is performed with a direct enhanced latching readout, as described in \citet{harvey-collard2018}, and using a single-electron transistor (SET) in series with a SiGe heterojunction bipolar transistor (HBT) cryoamplifier \cite{curry2015}. Triplet return probabilities $P(T)$ are calculated from the average of readout traces referenced to a known charge configuration to eliminate the slow charge sensor (CS) current fluctuations.

The external magnetic field $B_\text{ext}$ is applied in-plane along the [100] or [110] crystallographic orientations. The [100] orientation is used for all the experiments unless otherwise specified. The [110] orientation was obtained by rotating the sample in a separate cooldown. The device parameters (voltages, ST splittings, etc.) remained very similar between cooldowns, except for slight changes in the tunnel couplings.

\section{Results}

\begin{figure}[tbp]
   \centering
   \includegraphics{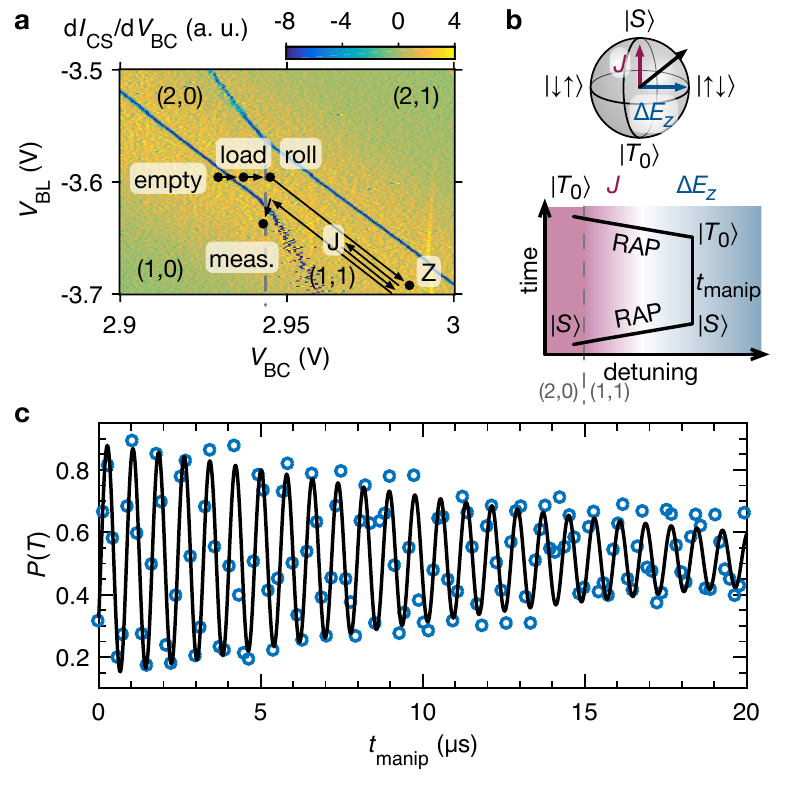} 
   \caption{\figletter{a} Charge stability diagram of the two-electron double QD. The typical pulse sequence steps are indicated. They consist of an emptying step (empty) where the charge occupancy is reset and the current referenced, a load step (load) for singlet or mixture preparation, a transient point (roll) to set the pulse trajectories and rates, some steps in $(1,1)$ (points Z and J) for spin control, a return to ``roll'' step, and an enhanced latching readout step (meas.). The roll point can be placed on either sides of the $S{-}T_-$ anticrossing depending on the goals. More details about the pulses can be found in the \supmat{} \SupRefSecSapRap{} and \SupRefSecPulseParams{}.    \figletter{b} ST qubit Bloch sphere and pulse sequence for $S{-}T_0$ rotations.     \figletter{c} Rotations between the $\ket{S}$ and $\ket{T_0}$ states versus the manipulation time $t_\text{manip}$. Those are enabled by a large $B_\text{ext} = 1 \tesla$ in the [100] direction, in spite of the relative absence of nuclear spins or magnetic materials. Optimal visibility is achieved in the rapid adiabatic passage (RAP) regime \cite{taylor2007,harvey-collard2017b} (see the \supmat{} \SupRefSecSapRap{}). The visibility is $73\pc$, limited largely by preparation and readout errors. The dephasing time is $15 \us$ for this single trace acquired in 4 minutes.}
   \label{fig:pulse}
\end{figure}

A charge stability diagram of the two-electron double QD and the typical location of the pulse sequence steps are shown in \figref{fig:pulse}. We use rotations between the $\ket{S}$ and $\ket{T_0} = (\ket{\uparrow\downarrow}+\ket{\downarrow\uparrow})/\sqrt{2}$ states to measure the difference in Zeeman energy perpendicular to the quantization axis $\DEz$ between the two QDs. These rotations appear with the application of an external magnetic field $B_\text{ext}$, as reported in \citet{jock2018}, in spite of the (relative) absence of lattice nuclear spins or magnetic materials. The inhomogeneous dephasing time saturates at $T_2^\star = 3.4 \pm 0.3 \us$ after 2~h of data averaging. This value is consistent with magnetic noise from residual \twentynineSi{} hyperfine coupling with the electron spins, and with other reported values \cite{witzel2010,assali2011,eng2015,rudolph2016,jock2018}.

\begin{figure}[tbp]
   \centering
   \includegraphics{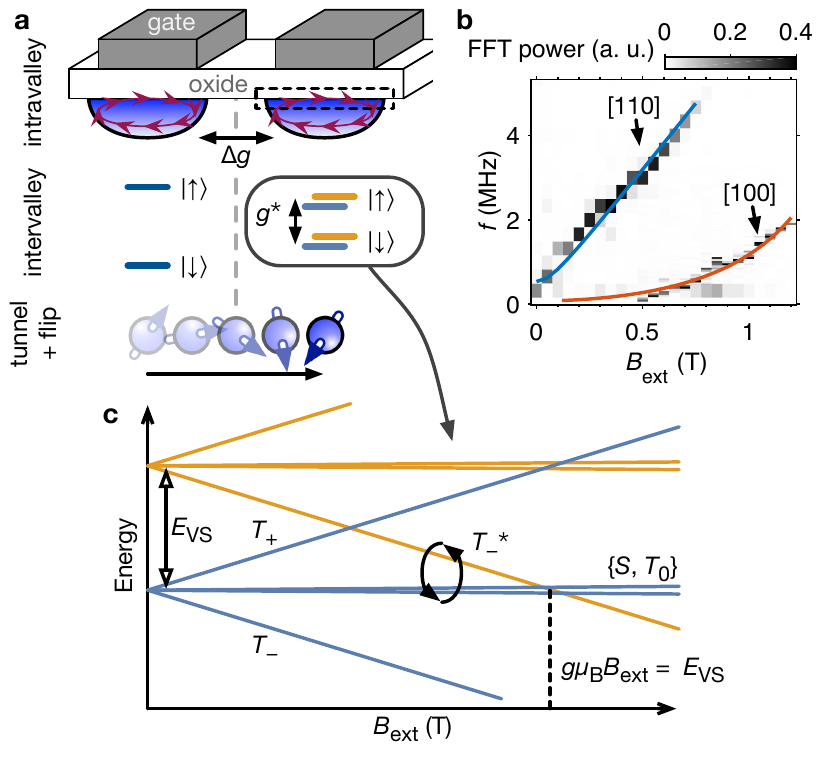} 
   \caption{\figletter{a} The three different spin-orbit effects for electrons in a silicon MOS nanodevice identified in this Letter. First, spin-orbit interaction in each dot leads to a renormalization of their effective $g$-factors. This entails an interaction of the form $\DEz = \Delta g \mu_\text{B} B_\text{ext}$ (see \figref{fig:pulse}b). Second, intervalley spin coupling can change the Landé factor $g^*$ of one dot in particular, leading to an interaction of the form $\DEz \propto B_\text{ext}^2/\of{E_\text{VS}-g \mu_\text{B} B_\text{ext}}$. Third, electron motion during a tunneling event can induce a spin flip that couples the $(2,0)S$ and $(1,1)T_-$ states, as shown also in \figref{fig:funnel}.         \figletter{b} Fast Fourier transform (FFT) power of the $S{-}T_0$ rotations versus $B_\text{ext}$. For the [110] field orientation, the linear interdot effect dominates. For the [100] orientation, the linear interdot effect is suppressed; however, a second-order effect consistent with an intervalley mechanism is observed. The solid lines are fit to a complete Hamiltonian model detailed in the \supmat{} \SupRefSecModel{} and agree well with the simple analytical forms described above.    \figletter{c} The intervalley spin coupling in one of the dots perturbs the $S{-}T_0$ energy difference at the second order through the excited valley $\ket{T_-^*}$ state. This simple model neglects the excited valley state of the other dot, which is higher in energy.}
   \label{fig:freqdep}
\end{figure}

To investigate the physical origin of the $S{-}T_0$ rotations, we vary the strength of $B_\text{ext}$ along two orientations measured in successive cooldowns. We identify different dominant spin-orbit mechanisms for these two orientations, with the results summarized in \figref{fig:freqdep}.

\begin{figure}[tbp]
   \centering
   \includegraphics{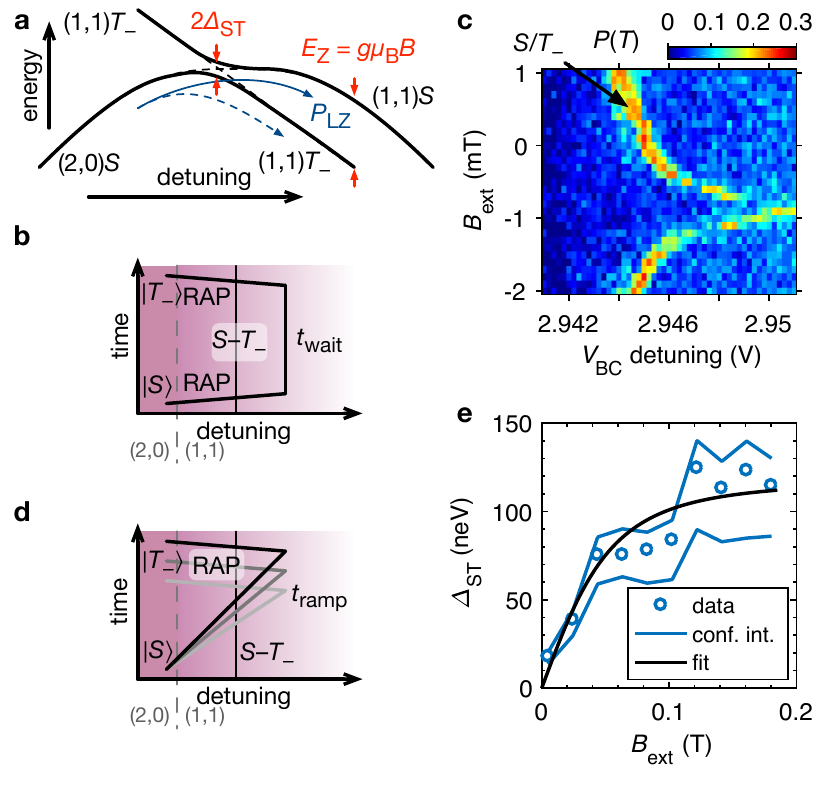} 
   \caption{\figletter{a} Anticrossing of the $\ket{S}$ and $\ket{T_-}$ states. The separation $E_\text{Z} = g \mu_\text{B} B$ between the $(1,1)S$ and $(1,1)T_-$ states is tuned with the magnetic field $B$. The coupling strength $\varDelta_\text{ST}$ can come from both a tunneling plus spin-flip mechanism that couples $(1,1)$ and $(2,0)$ states, and from the hyperfine interaction with lattice nuclear spins which couples $(1,1)$ states together. For our residual \twentynineSi{} concentration, the latter should be less than $0.6 \neV$ \cite{witzel2010,assali2011}.     \figletter{b} Pulse sequence for the spin funnel measurement. When the detuning pulse falls on the transition, mixing between $\ket{S}$ and $\ket{T_-}$ occurs.     \figletter{c} Spin funnel with $t_\text{wait} = 20 \us$. This measurement allows us to calibrate the tunnel coupling $t_\text{c}$ and energy ramp rate $\D E/\D t$ of LZ sweeps.     \figletter{d} Pulse sequence to probe the $S{-}T_-$ gap through LZ transition probabilities.    \figletter{e} The magnetic field dependence of the gap is fit to a simple model that includes a constant spin-orbit term $\varOmega_\text{SO}$ and charge hybridization, see the \supmat{} \SupEqnOmegaSOfit{}. The confidence interval is explained in the \supmat{} \SupRefSecSTMinus{}. The field is in the [100] direction.}
   \label{fig:funnel}
\end{figure}

The first mechanism is a first-order intravalley effect observed both in this device and in \citet{jock2018}. The Zeeman drive is a difference in effective Landé $g$-factor $\Delta g$ between the two QDs:
\ma{
	\DEz = \Delta g \mu_\text{B} B_\text{ext}
	. \label{eq:firstorderterm}
}
This effect dominates in the [110] field orientation. It is not predicted to depend on the double QD orientation, as shown by the different positions for the two QDs in this Letter compared with \citet{jock2018}.

The second mechanism, newly reported here, is consistent with an intervalley spin-orbit interaction \cite{yang2013b,hao2014,corna2018}. The smaller and nonlinear behavior versus magnetic field from \figref{fig:freqdep}(b) suggests a second-order interaction with an excited valley $\ket{T_-^*}$ state, as shown in \figref{fig:freqdep}(c). For simplicity, we consider only the QD with the lowest valley splitting $E_\text{VS}$. In the other QD, this interaction is suppressed by the larger $E_\text{VS}$. Using perturbation theory, we have 
\ml{
	E_S-E_{T_0} = \of{ E_S^{(0)}-E_{T_0}^{(0)} } + \ldots \\
	\of{ \frac{\abs{\bigbra{T_-^{*(0)}}{H_\text{SO}}\bigket{S^{(0)}}}^2}{E_S^{(0)}-E_{T_-^*}^{(0)}}
	-\frac{\abs{\bigbra{T_-^{*(0)}}{H_\text{SO}}\bigket{{T_0}^{(0)}}}^2}{E_{T_0}^{(0)}-E_{T_-^*}^{(0)}} }
	. \label{eq:perturbation1}
}
Here, $E_\psi^{(i)}$ is the energy of the state $\ket{\psi}$ at the $i^\text{th}$ order, and $H_\text{SO}$ is the spin-orbit interaction Hamiltonian. We note that the first term on the right-hand side is the effect of \eqnref{eq:firstorderterm}. This first term is largely suppressed for the [100] field orientation, as in \citet{jock2018}. The second term on the right-hand side can be simplified as follows. The matrix elements $\bigbra{T_-^{*(0)}}{H_\text{SO}}\bigket{S^{(0)}}$ and $\bigbra{T_-^{*(0)}}{H_\text{SO}}\bigket{{T_0}^{(0)}}$ are both proportional to $B_\text{ext}$, as explained in the \supmat{} \SupRefSecModel{}. Therefore, \eqnref{eq:perturbation1} simplifies to
\ma{
	\DEz = \frac{\bar\beta_C^2 B_\text{ext}^2}{2(E_\text{VS}-g \mu_\text{B} B_\text{ext})}
	. \label{eq:secondorderterm}
}
Here, $\bar\beta_C$ is a measure of the Dresselhaus spin coupling of the C QD. The above treatment is simplistic but provides intuition about the physical mechanism and agrees well with the more detailed analysis of the \supmat{} \SupRefSecModel{}. We extract a value of $\bar\beta_C = 0.7 \ueVpT$ for the experimental data in \figref{fig:freqdep}(b). While this value is in qualitative agreement with previously inferred values for single spins \cite{yang2013b,hao2014,corna2018}, the experimental agreement with the model in \figref{fig:freqdep}(b) isn't perfect. As demonstrated in Fig.\ 6(c) of \citet{harvey-collard2017b}, we have also observed a detuning dependence of the rotation frequency (and hence $\bar\beta_C$). This suggests that $\bar\beta_C$ depends on the detuning via the microscopic details of the electron confinement and/or the electric field.

The ST qubit allows us to probe a third spin-orbit effect that involves a tunneling plus spin-flip mechanism, newly reported here for a silicon device. We apply the method featured in \citet{nichol2015a} to measure the $S{-}T_-$ gap $\varDelta_\text{ST}$. This method consists of mapping the position of the $S{-}T_-$ anticrossing using the spin funnel technique \cite{petta2005} and probing the gap size using Landau-Zener-Stückelberg-Majorana (LZ) transitions \cite{shevchenko2010}. The pulse sequence and results are shown in \figref{fig:funnel}. The data analysis is explained in the \supmat{} \SupRefSecSTMinus{}.
We find $\varDelta_\text{ST} = 113 \pm 22 \neV$. This gap is expected to depend upon the orientation of both the magnetic field as well as the axis of the double QD (through the electron motion). From this value, we can estimate a spin-orbit length $\lambda_\text{SO} \approx 1 \um$, which is slightly smaller than the bulk value for GaAs and 20 times smaller than the bulk Si value \cite{mehl2014a}. Therefore, the spin-orbit interaction in this silicon nanoscale device is comparable to the bulk value observed in larger spin-orbit materials.

Finally, we report in the \supmat{} \SupRefSecRelax{} measurements of the ST qubit relaxation time, and discuss its potential relation to the spin-orbit effects discussed here. We also explore in \supmat{} \SupRefSecDNP{} the possibility to use a DDNP sequence to enhance the coherence of the qubit and induce hyperfine-driven rotations despite the isotopic enrichment.
\footnote{See \supmat{} for additional details about slow and rapid adiabatic passage, pulse sequence details, measurement of the $S{-}T_-$ gap, relaxation time, dynamic nuclear polarization, and spin-orbit interaction model, which includes Refs.\ \cite{zener1932,fasth2007,dial2013}.}

\section{Conclusion}

In summary, we report three different spin-orbit effects for electrons in an isotopically enriched silicon double QD device. We observe both coherent $S{-}T_0$ rotations and incoherent $S{-}T_-$ mixing that are consistent with a spin-orbit interaction much larger than bulk silicon values. We extend an analytical theory based on broken crystal symmetries at the silicon--dielectric interface that captures first- and second-order effects. Based on this theory and the results by \citet{jock2018}, we predict that the two $S{-}T_0$ effects could be eliminated with an out-of-plane magnetic field orientation since the dot-localized electron momentum at the interface vanishes in total. The $S{-}T_-$ effect could potentially persist in such an orientation due to the interdot electron tunneling.
Our results have implications for a variety of spin qubit encodings, like the $S{-}T_0$, the $S{-}T_-$ \cite{qi2017a} and the spin-1/2 qubits, for extending the coherence of ST silicon qubits through a DDNP, for single-spin control and relaxation, and for two-qubit coupling schemes based on the exchange interaction. For example, exchange-based two-qubit gates are in many ways operations similar to those in a ST qubit \cite{bonesteel2001,milivojevic2018,klinovaja2012,veldhorst2014a}. Beyond qubits, our results help understand additional spin-orbit effects that emerge in nanostructures.

\section*{Acknowledgements}
The authors thank David S.\ Simons and Joshua M.\ Pomeroy from National Institute of Standards and Technology for assistance with secondary ion mass spectrometry measurements on the \twentyeightSi{} epitaxial layer.
The authors recently became aware of simultaneous work by \citet{tanttu2018} that covers related topics.
This work was performed, in part, at the Center for Integrated Nanotechnologies, an Office of Science User Facility operated for the U.S.\ Department of Energy (DOE) Office of Science. Sandia National Laboratories is a multimission laboratory managed and operated by National Technology and Engineering Solutions of Sandia, LLC, a wholly owned subsidiary of Honeywell International, Inc., for the DOE's National Nuclear Security Administration under contract \mbox{DE-NA0003525}. This paper describes objective technical results and analysis. Any subjective views or opinions that might be expressed in the paper do not necessarily represent the views of the U.S.\ DOE or the United States Government.

\section*{Author contributions}
P.H.-C., R.M.J.\ and M.S.C.\ designed the experiments. 
P.H.-C.\ performed the experiments and analyzed the results.
N.T.J.\ developed the theory with help from V.S.\ and A.M.M.
Furthermore, C.B.-O.\ performed experiments in the [110] field orientation.
\mbox{P.H.-C.}, N.T.J., C.B.-O., R.M.J., V.S., A.M.M.\ and M.S.C.\ discussed the results.
R.M.J.\ performed experiments on another device that establishes the reproducibility of some results.
D.R.W., J.M.A., R.P.M., J.R.W., T.P.\ and M.S.C.\ designed the process flow and fabricated the devices. 
M.L.\ provided the experimental setup for the work.
M.P.-L.\ and D.R.L.\ helped develop the project and provided counsel.
M.S.C.\ supervised the combined effort, including coordinating fabrication and identifying modeling needs. P.H.-C., N.T.J.\ and M.S.C.\ wrote the manuscript with input from all coauthors.

\bibliographystyle{apsrev4-1-title} 
\bibliography{/Users/Patrick/OneDrive/Papers/PHC-cloud}


\makeatletter
\renewcommand\thesection{\mbox{S\arabic{section}}}
\makeatother
\setcounter{section}{0}     

\newcounter{supfigure} \setcounter{supfigure}{0} 
\makeatletter
\renewcommand\thefigure{\mbox{S\arabic{supfigure}}}
\makeatother

\newcounter{suptable} \setcounter{suptable}{0} 
\makeatletter
\renewcommand\thetable{\mbox{S\arabic{suptable}}}
\makeatother

\newenvironment{supfigure}[1][]{\begin{figure}[#1]\addtocounter{supfigure}{1}}{\end{figure}}
\newenvironment{supfigure*}[1][]{\begin{figure*}[#1]\addtocounter{supfigure}{1}}{\end{figure*}}
\newenvironment{suptable}[1][]{\begin{table}[#1]\addtocounter{suptable}{1}}{\end{table}}
\newenvironment{suptable*}[1][]{\begin{table*}[#1]\addtocounter{suptable}{1}}{\end{table*}}

\makeatletter
\renewcommand\theequation{S\arabic{equation}}
\renewcommand\theHequation{S\arabic{equation}} 
\makeatother
\setcounter{equation}{0}

\newcommand{\RefFigPulse}{\figref{fig:pulse}} 
\newcommand{\RefFigFreqDep}{\figref{fig:freqdep}} 
\newcommand{\RefFigFunnel}{\figref{fig:funnel}} 
\newcommand{\MainTextEqnSecondorderterm}{\eqnref{eq:secondorderterm}} 

\onecolumngrid 
\clearpage
{\centering
\large\textbf
{Supplementary information for: \\ \mytitle} \\ \rule{0pt}{12pt}
}
\twocolumngrid


\section{Slow and rapid adiabatic passage}
\label{sec:saprap}

In \figref{fig:saprap}, we demonstrate two ways to shuttle spins in a ST qubit \cite{taylor2007,harvey-collard2017b}. In the slow adiabatic passage (SAP) regime, the initial state is slowly (i.e., adiabatically with respect to both spin and charge) mapped from the $\{\ket{S},\ket{T_0}\}$ eigenbasis to the $\{ \ket{\uparrow\downarrow}, \ket{\downarrow\uparrow} \}$ eigenbasis. In the rapid adiabatic passage (RAP) regime, the initial state is left mostly unchanged (i.e., diabatically with respect to spin and adiabatically for charge) between eigenbases.
\begin{supfigure}[htbp]
   \centering
   \includegraphics{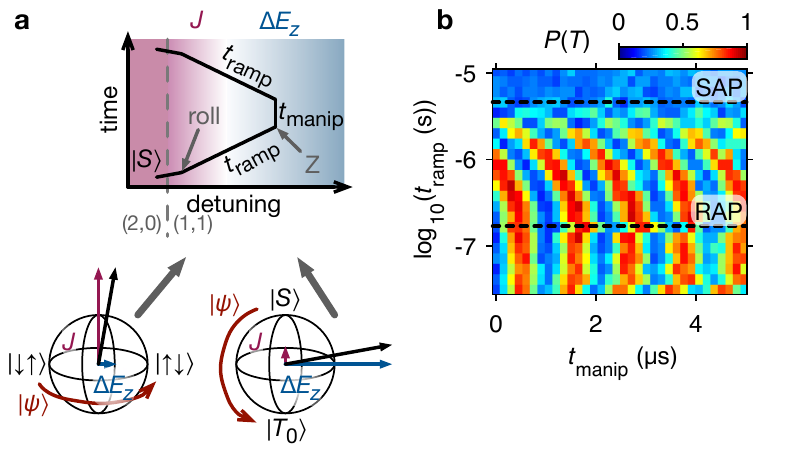} 
   \caption{\figletter{a} Pulse sequence to probe the SAP and RAP regimes.     \figletter{b} Experimental result. For slow ramps, the adiabatic mapping of spin means that the initial state, a singlet, is preserved after the sequence. For faster ramps, the state barely evolves during this ramp, resulting in time evolution of the state during the manipulation time. For the fastest ramps, charge adiabaticity is no longer preserved, resulting in a loss of visibility.}
   \label{fig:saprap}
\end{supfigure}

\section{Pulse sequence details}
\label{sec:pulseparams}
The alternative current (AC) component of pulses in the experiment is applied using a Tektronix AWG7122C arbitrary waveform generator with two synchronized channels for the BC and BL gates. The waveform is composed of direct current (DC) and AC components, and applied to the gates through a resistive-capacitive bias tee on the cold printed circuit board. The waveforms are applied such that all target points are fixed in the charge stability diagram, except the ones explicitly varied for a particular measurement (e.g.\ manipulation time or position of point Z). An example of pulse sequence parameters is given in \tabref{tab:pulseparams}.
\begin{suptable}
\caption{Table of pulse sequence points (as defined in main text \RefFigPulse{}a), ramp time to point (from previous point), and wait time at point for a pulse sequence example that probes spin rotations between the $\{\ket{S},\ket{T_0}\}$ states. The sequence is played in a loop.}
\begin{center}
\begin{tabular}{l|r|r|c}
	\hline\hline
	Point	& Ramp time ($\upmu$s)	& Wait time ($\upmu$s)	& State after wait\\
	\hline
	empty	& 1					& 10 				& $(1,0)$\\
	load		& 0					& 60 				& $(2,0)S$\\
	roll		& 0.1				& 0 					& $(1,1)S$\\
	Z		& 0.3				& 0 through 10 		& $(1,1)S \leftrightarrow (1,1)T_0$\\
	roll		& 0.3				& 0 					& $(1,1)S$ or $(1,1)T_0$\\
	meas.	& 1					& 100 				& $(1,0)$ or $(1,1)T_0$\\
	\hline\hline
\end{tabular}
\end{center}
\label{tab:pulseparams}
\end{suptable}

\section{Measurement of the \texorpdfstring{$\boldsymbol{S{-}T_-}$}{S-T-} gap}
\label{sec:stminussup}

We apply the technique featured in \citet{nichol2015a} to measure the $S{-}T_-$ gap, $\varDelta_\text{ST}$. The method consists of mapping the position of the $S{-}T_-$ anticrossing using the spin funnel technique \cite{petta2005}. The funnel allows to extract the tunnel coupling through the Zeeman energy. This is shown in the main text \RefFigFunnel{}. We find a half-gap tunnel coupling $t_\text{c} = 5.15 \pm 0.25 \ueV$.

\begin{supfigure}[htbp]
   \centering
   \includegraphics{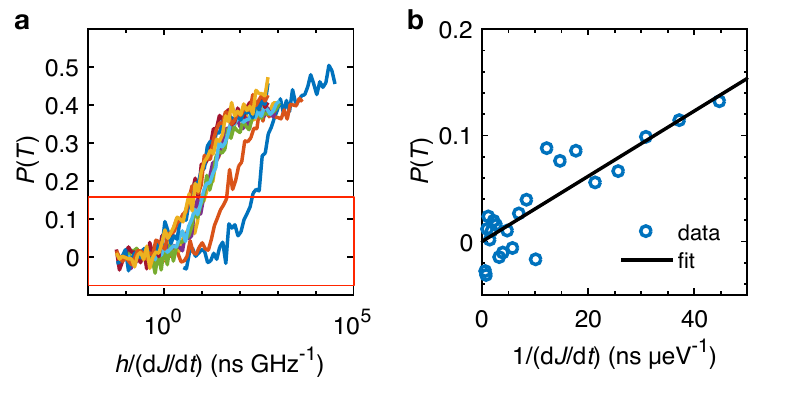} 
   \caption{\figletter{a} LZ sweep results for various values of $B_\text{ext}$. The energy ramp rate $\D J/\D t$ is calibrated using the experimentally-measured spin funnel.     \figletter{b} The gap $\varDelta_\text{ST}$ is extracted by fitting the previous data for $P(T) < 0.15$ using \eqnref{eq:poftLZ}. This example is for $B_\text{ext} = 5 \mT$.}
   \label{fig:sogap}
\end{supfigure}

The gap $\varDelta_\text{ST}$ can be probed using Landau-Zener-Stückelberg-Majorana (LZ) transitions \cite{shevchenko2010}. The pulse sequence is shown in the main text \RefFigFunnel{}. It consists of using a varying ramp rate through the $S{-}T_-$ transition, followed by a diabatic ramp back into $(2,0)$. The LZ probability of staying in the same state is $P_\text{LZ} = \Exp{-(2\pi\varDelta_\text{ST})^2/(h \D E/\D t)}$, where $\D E/\D t$ is the energy ramp rate evaluated at the $S{-}T_-$ anticrossing \cite{zener1932}. We note that $\D E/\D t = \D J/\D t$, with $J$ the exchange interaction measured from the spin funnel. After the pulse, $P(T) = 1-P_\text{LZ}$. In \figref{fig:sogap}(b), we show the result of such pulses for various values of $B_\text{ext}$. The $P(T)$ saturates close to $0.5$ rather than $1$. This has been attributed to charge noise by \citet{nichol2015a}. To mitigate the impact of this on the gap extraction procedure, we fit only the values for which $P(T) < 0.15$. The formula above can be simplified to
\ma{
	P(T) = \frac{(2\pi\varDelta_\text{ST})^2}{h \D E/\D t} .
	\label{eq:poftLZ}
}
This formula is used to extract the gap in \figref{fig:sogap}(b).

We plot the values obtained for $\varDelta_\text{ST}$ against the magnetic field in main text \RefFigFunnel{}(e). We find that the largest source of error is the probability calibration. The confidence interval is obtained by repeating the fit procedure using extremal values for the probability. The result is a moderate error in the scale of $\varDelta_\text{ST}$ that is consistent throughout the range, and hence it does not qualitatively affect the result. The resulting data is fit to a simple model that includes a constant spin-orbit term $\varOmega_\text{SO}$ and charge hybridization \cite{nichol2015a},
\ma{
	\varDelta_\text{ST} = \varOmega_\text{SO}\sin\Theta\sin\xi ,
	\label{eq:omegasofit}
}
where $\Theta = \arctan(B_\text{ext}/t_\text{c})$ and $\xi$ is the double QD angle with respect to crystallographic axes. We obtain a value of $\varOmega_\text{SO}\sin\xi = 113 \pm 22 \neV$. Due to the absence of a vector magnet, it was not possible to obtain the full angular dependence. The origin of the angle $\xi$ is therefore undetermined \cite{rancic2014}. In our analysis, we neglect the effect of residual \twentynineSi{} spins, because these are expected to contribute less than a nanoelectronvolt to this gap \cite{witzel2010,assali2011}. Experimentally, we can bound the hyperfine contribution to less than $5 \neV$ using a formula as in \citet{nichol2015a}.

We can use the relation $\lambda_\text{SO} = (t_\text{c} d)/(\sqrt{2}\varOmega_\text{SO})$, where $d$ is half of the interdot spacing, to estimate a spin-orbit length \cite{fasth2007,nichol2015a}. Using $d \approx 30 \nm$, and neglecting the angular factor for $\varOmega_\text{SO}\sin\xi$, we find $\lambda_\text{SO} \approx 1 \um$. 

\section{Relaxation time}
\label{sec:trelax}

The relaxation and excitation time $T_1$ of the $S{-}T_0$ qubit was measured both for this device and the one of \citet{jock2018} using SAP preparation and readout (as described in \figref{fig:saprap} but varying $t_\text{manip}$). We obtain values in the range of 30 to $100 \us$ in the regime where exchange is suppressed. These $T_1$ values are consistent with measured Hahn spin echo $T_2$ that seem limited by $T_1$. For the device featured in the main text, $T_1$ becomes larger as exchange is turned on, see \figref{fig:trelax}. This suggests that the relaxation and excitation mechanism limiting $T_1$ is suppressed when $J \gg \DEz$. These results contrast with measurements by \citet[see supplement]{dial2013} that show $T_1$ increasing to milliseconds when $J \ll \DEz$. The exact mechanism remains unclear at the moment; however, we note that the relaxation and excitation seems limited to the $\{\ket{S},\ket{T_0}\}$ subspace, as opposed to single-spin relaxation leading to a $\ket{T_-}$ ground state. This hints at a fluctuation of the quantization axis $\DEz$ itself, hypothetically due to charge noise or other electric fluctuations, which would couple through the microscopic details of the spin-orbit interaction at the interface. This would be consistent with the suppression of this decoherence when the quantization axis is dominated by $J$.

\begin{supfigure}[htbp]
   \centering
   \includegraphics{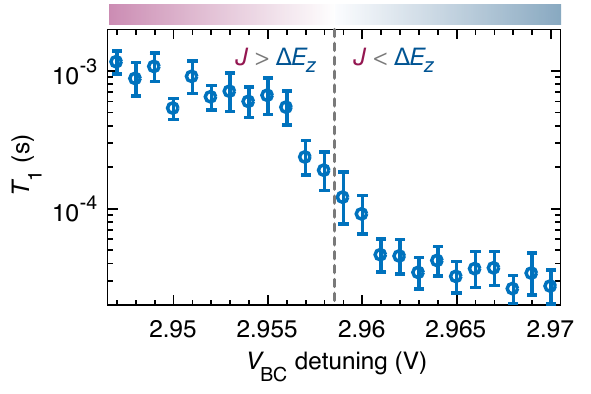} 
   \caption{Relaxation time $T_1$ as a function of the detuning in the $(1,1)$ charge configuration. The point where $J = \DEz$ (dashed line) is inferred from coherent manipulation experiments.}
   \label{fig:trelax}
\end{supfigure}

\section{Dynamic nuclear polarization}
\label{sec:dnp}

As a complementary experiment, we have looked for signatures of DDNP. This effect has been used in GaAs devices to prolong the spin coherence time \cite{bluhm2010} and induce a $\DEz$ for qubit control \cite{foletti2009}. It was shown that spin-orbit interaction can quench the ability to perform DDNP in double-QD devices \cite{nichol2015a}. 

For our silicon enrichment level, we expect that the maximum polarization achievable by flipping all nuclear spins in opposite ways is approximately $1 \MHz$ \cite{assali2011}. This is an impractical extremal scenario. We instead expect that the polarization could reach fractions of this value in a steady-state pump-probe experiment. The pump-probe experiment consists of one or many cycles of pseudo-SAP ramps through the $S{-}T_-$ anticrossing to ``pump'' the nuclear spins, followed by a probe cycle where the frequency of potential $S{-}T_0$ rotations is measured by varying the rotation time. The continuous repetitions should result in a steady-state polarization that could (i) enhance the qubit coherence time by slowing down the $S{-}T_0$ mixing, and/or (ii) result in a non-vanishing average $S{-}T_0$ rotation frequency.

Here we list some of the parameters used in our trials. We used two fields of $0.4 \mT$ and $150 \mT$. We interleaved one, two and three pump cycles with the probe cycle, and compared the results with those without pump cycles. The data was averaged in each case for periods of up to 6 hours of continuous pump-probing. A few ramp rates were tried, including one that aims at a moderate $P(T) \approx 0.3$ to avoid long incoherent mixing (pseudo-SAP).

In none of the results did we find meaningful differences between the pump and no-pump cases. This is not surprising, in light of the work of \citet{nichol2015a}, and suggests a spin-orbit origin of the $S{-}T_-$ mixing and quenching of DDNP by spin-orbit interaction.

\onecolumngrid
\section{Spin-orbit interaction model}
\label{sec:model}

\newcommand{\e}[1]{\ensuremath{\times 10^{#1}}}
\newcommand{\units}[1]{\ensuremath{\mathrm{#1}}}
\renewcommand{\bra}[1]{ \langle #1 \vert }
\renewcommand{\ket}[1]{ |{#1} \rangle }
\renewcommand{\braket}[2]{ \langle #1 \vert #2 \rangle }
\newcommand{\mtxelem}[3]{ \langle #1 \vert #2 \vert #3 \rangle }
\newcommand{\Ident}{\mbox{\ensuremath{1\hspace{-1mm}{\bf l}}}} 

\newcommand{\opS}{\operatorname{S}}
\newcommand{\opT}{\operatorname{T_0}}
\newcommand{\opTp}{\operatorname{T_+}}
\newcommand{\opTm}{\operatorname{T_-}}
\newcommand{\opST}{\operatorname{S-T_0}}
\newcommand{\TTwoStar}{$T_{2}^{*}$}

\newcommand{\gmb}{\ensuremath{g\sym{\mu}{B}}}
\newcommand{\comment}[1]{\textcolor{red}{#1}}


\def\simlt{\mathrel{\lower .3ex \rlap{$\sim$}\raise .5ex \hbox{$<$}}}
\def\simgt{\mathrel{\lower .3ex \rlap{$\sim$}\raise .5ex \hbox{$>$}}}

%
\onecolumngrid

Consider sitting at an interdot detuning $\epsilon$ that is well within the $(1,1)$ charge configuration, wherein $\epsilon$ is much larger than the interdot tunnel coupling $t$, $\epsilon \gg t$. In this case, we can approximate the two-electron states as being composed of the single-particle eigenstates of either the left (L) or right (R) quantum dots. Suppose, for now, that the valley splitting in the left dot, $\Delta_{\mathrm{vs},L}$ is significantly larger than for the right dot, $\Delta_{\mathrm{vs},R}$. As a consequence, the relevant low-energy excited state we ought to include is the excited valley state of the second dot, $\ket{\phi_{R^{*}}}$. In the following, we take the convention that the spin configuration $\ket{\uparrow}$ is oriented along the $\hat{z}$ crystallographic axis normal to the two-dimensional electron gas plane. The relevant $(1,1)$ two-electron states involving the ground valley states, $\ket{\phi_{L}}$, $\ket{\phi_{R}}$ are the following:
\begin{eqnarray}
\ket{S(1,1)} & = & \frac{1}{2} \big( \ket{\phi_{L} \phi_{R}} + \ket{\phi_{R} \phi_{L}} \big) \big( \ket{\uparrow \downarrow} - \ket{\downarrow \uparrow} \big) \\
\ket{T_{0}(1,1)} & = & \frac{1}{2} \big( \ket{\phi_{L} \phi_{R}} - \ket{\phi_{R} \phi_{L}} \big) \big( \ket{\uparrow \downarrow} + \ket{\downarrow \uparrow} \big)  \\
\ket{T_{+}(1,1)} & = & \frac{1}{\sqrt{2}} \big( \ket{\phi_{L} \phi_{R}} - \ket{\phi_{R} \phi_{L}} \big) \ket{\uparrow \uparrow}  \\
\ket{T_{-}(1,1)} & = & \frac{1}{\sqrt{2}} \big( \ket{\phi_{L} \phi_{R}} - \ket{\phi_{R} \phi_{L}} \big) \ket{\downarrow \downarrow} ,
\end{eqnarray}
while the relevant (1,1) two-electron states that involve the excited single-particle valley state $\ket{\phi_{R^{*}}}$ are
\begin{eqnarray}
\ket{S^{*}(1,1)} & = & \frac{1}{2} \big( \ket{\phi_{L} \phi_{R^{*}}} + \ket{\phi_{R^{*}} \phi_{L}} \big) \big( \ket{\uparrow \downarrow} - \ket{\downarrow \uparrow} \big)  \\
\ket{T_{0}^{*}(1,1)} & = & \frac{1}{2} \big( \ket{\phi_{L} \phi_{R^{*}}} - \ket{\phi_{R^{*}} \phi_{L}} \big) \big( \ket{\uparrow \downarrow} + \ket{\downarrow \uparrow} \big)  \\
\ket{T_{+}^{*}(1,1)} & = & \frac{1}{\sqrt{2}} \big( \ket{\phi_{L} \phi_{R^{*}}} - \ket{\phi_{R^{*}} \phi_{L}} \big) \ket{\uparrow \uparrow}  \\
\ket{T_{-}^{*}(1,1)} & = & \frac{1}{\sqrt{2}} \big( \ket{\phi_{L} \phi_{R^{*}}} - \ket{\phi_{R^{*}} \phi_{L}} \big) \ket{\downarrow \downarrow} .
\end{eqnarray}

We choose Cartesian coordinates $x, y, z$ defined along the crystallographic directions $[100]$, $[010]$ and $[001]$, respectively, and with $[001]$ perpendicular to the interface. This is shown in \figref{fig:model}. Let the applied magnetic field be given by
\begin{equation}
\mathbf{B} = (B_{x}, B_{y}, B_{z}) = B \big( \sin(\theta) \cos(\varphi), \sin(\theta) \sin(\varphi), \cos(\theta) \big).
\end{equation}

\begin{supfigure}[bp]
   \centering
   \includegraphics{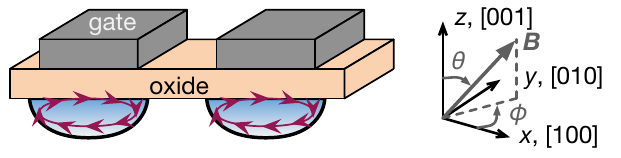} 
   \caption{Model coordinates system. The double QD axis is oriented close to the [110] crystallographic axis.}
   \label{fig:model}
\end{supfigure}

The full $8\times8$ Hamiltonian describing the $(1,1)$ charge sector is given by the following, in the basis
\begin{equation}
\mathcal{H}_{(1,1)} = \mathrm{span}\lbrace S(1,1), T_{0}(1,1), T_{+}(1,1), T_{-}(1,1), S^{*}(1,1), T_{0}^{*}(1,1), T_{+}^{*}(1,1), T_{-}^{*}(1,1) \rbrace ,
\label{eq:SpinBasisZ}
\end{equation}
and
\begin{equation}
H = \left(
\begin{array}{cc}
H_{\mathrm{GG}} & H_{\mathrm{GE}} \\
H_{\mathrm{GE}}^{\dagger} & H_{\mathrm{EE}}
\end{array}
\right),
\label{eq:Ham8x8}
\end{equation}
where
\begin{equation}
H_{\mathrm{GG}} =
\left(
\begin{array}{cccc}
0 & 0 & -\frac{1}{\sqrt{2}}(h_{\downarrow \uparrow}^{LL} - h_{\downarrow \uparrow}^{RR}) & \frac{1}{\sqrt{2}}(h_{\uparrow \downarrow}^{LL} - h_{\uparrow \downarrow}^{RR}) \\
\cdot & J_{11} & \frac{1}{\sqrt{2}}(h_{\downarrow \uparrow}^{LL} + h_{\downarrow \uparrow}^{RR})+\frac{1}{\sqrt{2}} g \mu_{B}(B_{x} + i B_{y}) & \frac{1}{\sqrt{2}}(h_{\uparrow \downarrow}^{LL} + h_{\uparrow \downarrow}^{RR})+\frac{1}{\sqrt{2}} g \mu_{B}(B_{x} - i B_{y}) \\
\cdot & \cdot & J_{11} + g \mu_{B} B_{z} & 0 \\
\cdot & \cdot & \cdot & J_{11} - g \mu_{B} B_{z}
\end{array}
\right) ,
\end{equation}

\begin{equation}
H_{\mathrm{GE}} =
\left(
\begin{array}{cccc}
0 & 0 & \frac{1}{\sqrt{2}} h_{\downarrow \uparrow}^{R R^{*}} & -\frac{1}{\sqrt{2}} h_{\uparrow \downarrow}^{R R^{*}} \\
0 & 0 & \frac{1}{\sqrt{2}} h_{\downarrow \uparrow}^{R R^{*}} & \frac{1}{\sqrt{2}} h_{\uparrow \downarrow}^{R R^{*}} \\
\frac{1}{\sqrt{2}} h_{\uparrow \downarrow}^{R R^{*}} & -\frac{1}{\sqrt{2}} h_{\uparrow \downarrow}^{R R^{*}} & 0 & 0 \\
-\frac{1}{\sqrt{2}} h_{\downarrow \uparrow}^{R R^{*}} & \frac{1}{\sqrt{2}} h_{\downarrow \uparrow}^{R R^{*}} & 0 & 0
\end{array}
\right) ,
\end{equation}

\begin{equation}
H_{\mathrm{EE}} =
\left(
\begin{array}{cccc}
\Delta_{\mathrm{vs},R} & 0 & -\frac{1}{\sqrt{2}} (h_{\downarrow \uparrow}^{L L} - h_{\downarrow \uparrow}^{R^{*} R^{*}}) & \frac{1}{\sqrt{2}} (h_{\uparrow \downarrow}^{L L} - h_{\uparrow \downarrow}^{R^{*} R^{*}}) \\
\cdot & \Delta_{\mathrm{vs},R} + J_{11} & \frac{1}{\sqrt{2}} (h_{\downarrow \uparrow}^{L L} + h_{\downarrow \uparrow}^{R^{*} R^{*}}) + \frac{1}{\sqrt{2}} g \mu_{B}(B_{x} + i B_{y}) & \frac{1}{\sqrt{2}} (h_{\uparrow \downarrow}^{L L} + h_{\uparrow \downarrow}^{R^{*} R^{*}}) + \frac{1}{\sqrt{2}} g \mu_{B}(B_{x} - i B_{y})\\
\cdot & \cdot &  \Delta_{\mathrm{vs},R} + J_{11} + g \mu_{B} B_{z} & 0 \\
\cdot & \cdot & \cdot & \Delta_{\mathrm{vs},R} +J_{11} - g \mu_{B} B_{z}
\end{array}
\right) .
\end{equation}
Here $J_{11}$ parameterizes the residual exchange energy at the given operating point in the $(1,1)$ configuration, $\mu_{B} = 57.88 \ \mathrm{\mu eV\,T}^{-1}$ is the Bohr magneton, $g=2$ is the $g$-factor of bulk Si, and
\begin{equation}
h_{s_{1} s_{2}}^{MN} = \mtxelem{\phi_{M} s_{1}}{H_{\mathrm{SO}}}{\phi_{N} s_{2}}
\label{eq:SOMtxElem}
\end{equation}
with $s_{1}, s_{2} \in \lbrace \uparrow, \downarrow \rbrace$, $M, N \in \lbrace L, R, R^{*} \rbrace$. The spin-orbit coupling Hamiltonian $H_{\mathrm{SO}}$ is given by
\begin{eqnarray}
H_{\mathrm{SO}} & = & H_{R} + H_{D}  \\
& = & \delta(z) \big[ \gamma_{R} (P_{y}\sigma_{x} -P_{x} \sigma_{y}) + \gamma_{D} (P_{x} \sigma_{x} - P_{y} \sigma_{y}) \big],
\end{eqnarray}
with $z=0$ denoting the position of the Si/SiO$_{2}$ interface and $P_{x}, P_{y}$ the kinetic momenta along the $\hat{x}$ and $\hat{y}$ crystallographic axes.

We now turn our attention to evaluating the matrix elements \eqnref{eq:SOMtxElem}.
Since $\ket{\phi_{R}}$ and $\ket{\phi_{R^{*}}}$ are the ground and first excited valley states of the right dot, respectively, and occupy the two-dimensional $\pm z$ valley subspace, we can express them as
\begin{eqnarray}
\braket{\mathbf{r}}{\phi_{R}} & = & \frac{1}{\sqrt{2}} \big( e^{i k_{0} z} + e^{i \varphi_{v,\mathrm{R}}} e^{-i k_{0} z} \big) \psi_{R}(\mathbf{r}) \\
\braket{\mathbf{r}}{\phi_{R^{*}}} & = & \frac{1}{\sqrt{2}} \big( e^{i k_{0} z} - e^{i \varphi_{v,\mathrm{R}}} e^{-i k_{0} z} \big) \psi_{R}(\mathbf{r}),
\end{eqnarray}
where $\varphi_{v,\mathrm{R}}$ denotes a valley phase factor (relative complex phase between the $+z$ and $-z$ valley components) for the right dot and $\psi_{R}(\mathbf{r})$ is an envelope function that is common to these two lowest valley eigenstates. Similarly, for the left dot we have
\begin{equation}
\braket{\mathbf{r}}{\phi_{L}} = \frac{1}{\sqrt{2}} \big( e^{i k_{0} z} + e^{i \varphi_{v,\mathrm{L}}} e^{-i k_{0} z} \big) \psi_{L}(\mathbf{r}).
\end{equation}
Evaluating the interface-localized momentum matrix elements for these valley eigenstates along the lines of the analysis in \citet{jock2018}, we have
\begin{eqnarray}
\mtxelem{\phi_{L}}{\delta(z) P_{k}}{\phi_{L}} & = & (1 + \cos(\varphi_{v,\mathrm{L}})) \mtxelem{\psi_{L}}{\delta(z)P_{k}}{\psi_{L}} \\
\mtxelem{\phi_{R}}{\delta(z) P_{k}}{\phi_{R}} & = & (1 + \cos(\varphi_{v,\mathrm{R}})) \mtxelem{\psi_{R}}{\delta(z)P_{k}}{\psi_{R}} \\
\mtxelem{\phi_{R}}{\delta(z) P_{k}}{\phi_{R^{*}}} & = & -i \sin(\varphi_{v,\mathrm{R}}) \mtxelem{\psi_{R}}{\delta(z)P_{k}}{\psi_{R}},
\end{eqnarray}
with $k \in \lbrace x, y \rbrace$ and where $\mtxelem{\psi_{R}}{\delta(z)P_{x,y}}{\psi_{R}}$ are matrix elements that are proportional to the applied magnetic field. These latter factors also depend on lateral confinement and vertical electric field, wrapped into the parameters $\lambda^{x,y}$:
\begin{eqnarray}
\mtxelem{\psi_{L}}{\delta(z)P_{x}}{\psi_{L}} & = & \lambda_{L}^{x} B_{y} \\
\mtxelem{\psi_{L}}{\delta(z)P_{y}}{\psi_{L}} & = & -\lambda_{L}^{y} B_{x} \\
\mtxelem{\psi_{R}}{\delta(z)P_{x}}{\psi_{R}} & = & \lambda_{R}^{x} B_{y} \\
\mtxelem{\psi_{R}}{\delta(z)P_{y}}{\psi_{R}} & = & -\lambda_{R}^{y} B_{x}.
\end{eqnarray}
In the following, we will assume that the left and right dots are nearly symmetric, such that $\lambda_{L}^{x} = \lambda_{L}^{y} = \lambda_{L}$ and $\lambda_{R}^{x} = \lambda_{R}^{y} = \lambda_{R}$. Defining the Rashba and Dresselhaus spin-orbit (SO) coupling strengths for the left and right dots as
\begin{eqnarray}
\alpha^{L} = \gamma_{R} \lambda_{L} \\
\alpha^{R} = \gamma_{R} \lambda_{R} \\
\beta^{L} = \gamma_{D} \lambda_{L} \\
\beta^{R} = \gamma_{D} \lambda_{R},
\end{eqnarray}
we have 
\begin{eqnarray}
h_{\downarrow \uparrow}^{LL} & = & (1 + \cos(\varphi_{v,\mathrm{L}})) \big[ -\alpha^{L} (B_{x} + i B_{y}) + \beta^{L} (B_{y} + i B_{x})\big] \\
h_{\uparrow \downarrow}^{LL} & = & (1 + \cos(\varphi_{v,\mathrm{L}})) \big[ -\alpha^{L} (B_{x} - i B_{y}) + \beta^{L} (B_{y} - i B_{x})\big] \\
h_{\downarrow \uparrow}^{RR} & = & (1 + \cos(\varphi_{v,\mathrm{R}})) \big[ -\alpha^{R} (B_{x} + i B_{y}) + \beta^{R} (B_{y} + i B_{x})\big] \\
h_{\uparrow \downarrow}^{RR} & = & (1 + \cos(\varphi_{v,\mathrm{R}})) \big[ -\alpha^{R} (B_{x} - i B_{y}) + \beta^{R} (B_{y} - i B_{x})\big] \\
h_{\downarrow \uparrow}^{RR^{*}} & = & -i\sin(\varphi_{v,\mathrm{R}}) \big[ -\alpha^{R} (B_{x} + i B_{y}) + \beta^{R} (B_{y} + i B_{x})\big] \\
h_{\uparrow \downarrow}^{RR^{*}} & = & -i\sin(\varphi_{v,\mathrm{R}}) \big[ -\alpha^{R} (B_{x} - i B_{y}) + \beta^{R} (B_{y} - i B_{x})\big] \\
h_{\downarrow \uparrow}^{R^{*}R^{*}} & = & (1 - \cos(\varphi_{v,\mathrm{R}})) \big[ -\alpha^{R} (B_{x} + i B_{y}) + \beta^{R} (B_{y} + i B_{x})\big] \\
h_{\uparrow \downarrow}^{R^{*}R^{*}} & = & (1 - \cos(\varphi_{v,\mathrm{R}})) \big[ -\alpha^{R} (B_{x} - i B_{y}) + \beta^{R} (B_{y} - i B_{x})\big].
\end{eqnarray}
We now have explicit expressions for all matrix elements of the Hamiltonian \eqnref{eq:Ham8x8}. Note that we have shown that the intravalley and intervalley spin-orbit coupling (SOC) matrix elements all scale linearly with the applied magnetic field, following an extended version of the analysis previously detailed in \citet{jock2018}.

\subsection{Reduced-dimensional model}
Since we are interested primarily in the two-dimensional space spanned by the lowest-energy singlet and unpolarized triplet states, to arrive at an analytic expression for the ST rotation frequency we now evaluate perturbatively the action of these other six states on our qubit subspace through intravalley and intervalley spin-orbit coupling. First, we transform our basis from the original spin basis defined with respect to the $z$-axis, given in \eqnref{eq:SpinBasisZ}, into the spin basis defined by the applied magnetic field, $\mathbf{B}=B(\sin(\theta) \cos(\varphi), \sin(\theta) \sin(\varphi), \cos(\theta))$. In the following, we use a tilde to denote states associated with the spin basis determined by the applied magnetic field or operators defined in this basis. Diagonalizing the $2\times2$ Zeeman Hamiltonian to obtain the spin eigenstates associated with the applied magnetic field orientation,
\begin{equation}
H_{\mathrm{Zee}} = \left(
\begin{array}{cc}
\cos{\theta} & e^{-i \varphi} \sin{\theta} \\
e^{i \varphi} \sin{\theta} & -\cos{\theta}
\end{array}
\right),
\end{equation}
we have
\begin{eqnarray}
\ket{\tilde{\uparrow}} & = & e^{-i\varphi/2} \cos{(\theta/2)} \ket{\uparrow} + e^{i\varphi/2} \sin{(\theta/2)} \ket{\downarrow} \\
\ket{\tilde{\downarrow}} & = & -e^{-i\varphi/2} \sin{(\theta/2)} \ket{\uparrow} + e^{i\varphi/2} \cos{(\theta/2)} \ket{\downarrow}
\end{eqnarray}
and hence
\begin{eqnarray}
\ket{\tilde{T}_{-}(1,1)} & = & e^{-i \varphi} \sin^{2}\left(\frac{\theta}{2}\right) \ket{T_{+}(1,1)} + e^{i \varphi} \cos^{2}\left(\frac{\theta}{2}\right) \ket{T_{-}(1,1)} - \frac{1}{\sqrt{2}} \sin{\theta} \ket{T_{0}(1,1)} \nonumber \\
\ket{\tilde{S}(1,1)} & = & \ket{S(1,1)} \nonumber \\
\ket{\tilde{T}_{0}(1,1)} & = & \cos{\theta} \ket{T_{0}(1,1)} + \frac{1}{\sqrt{2}} \sin{\theta} \big( e^{i \varphi} \ket{T_{-}(1,1)} - e^{-i \varphi} \ket{T_{+}(1,1)} \big) \nonumber \\
\ket{\tilde{T}_{+}(1,1)} & = & e^{-i \varphi} \cos^{2}\left(\frac{\theta}{2}\right) \ket{T_{+}(1,1)} + e^{i \varphi} \sin^{2}\left(\frac{\theta}{2}\right) \ket{T_{-}(1,1)} + \frac{1}{\sqrt{2}} \sin{\theta} \ket{T_{0}(1,1)} \nonumber \\
\ket{\tilde{T}_{-}^{*}(1,1)} & = & e^{-i \varphi} \sin^{2}\left(\frac{\theta}{2}\right) \ket{T_{+}^{*}(1,1)} + e^{i \varphi} \cos^{2}\left(\frac{\theta}{2}\right) \ket{T_{-}^{*}(1,1)} - \frac{1}{\sqrt{2}} \sin{\theta} \ket{T_{0}^{*}(1,1)} \nonumber \\
\ket{\tilde{S}^{*}(1,1)} & = & \ket{S^{*}(1,1)} \nonumber \\
\ket{\tilde{T}_{0}^{*}(1,1)} & = & \cos{\theta} \ket{T_{0}^{*}(1,1)} + \frac{1}{\sqrt{2}} \sin{\theta} \big( e^{i \varphi} \ket{T_{-}^{*}(1,1)} - e^{-i \varphi} \ket{T_{+}^{*}(1,1)} \big) \nonumber \\
\ket{\tilde{T}_{+}^{*}(1,1)} & = & e^{-i \varphi} \cos^{2}\left(\frac{\theta}{2}\right) \ket{T_{+}^{*}(1,1)} + e^{i \varphi} \sin^{2}\left(\frac{\theta}{2}\right) \ket{T_{-}^{*}(1,1)} + \frac{1}{\sqrt{2}} \sin{\theta} \ket{T_{0}^{*}(1,1)} \label{eq:FullSpinBasis} .
\end{eqnarray}
Equivalently, the unitary transformation from the basis $\lbrace S(1,1), T_{0}(1,1), T_{+}(1,1), T_{-}(1,1)\rbrace$ to $\lbrace S(1,1), \tilde{T}_{0}(1,1), \tilde{T}_{+}(1,1), \tilde{T}_{-}(1,1)\rbrace$ is given by the unitary
\begin{equation}
U=\left(
\begin{array}{cccc}
1 & 0 & 0 & 0 \\
0 & \cos(\theta) & -\frac{1}{\sqrt{2}} \sin(\theta) e^{i \varphi} & \frac{1}{\sqrt{2}} \sin(\theta) e^{-i \varphi} \\
0 & \frac{1}{\sqrt{2}} \sin(\theta) & \cos^{2}(\theta/2) e^{i \varphi} & \sin^{2}(\theta/2) e^{-i \varphi} \\
0 & -\frac{1}{\sqrt{2}} \sin(\theta) & \sin^{2}(\theta/2) e^{i \varphi} & \cos^{2}(\theta/2) e^{-i \varphi}
\end{array}
\right) .
\label{eq:UnitaryU}
\end{equation}

Transforming the Hamiltonian \eqnref{eq:Ham8x8} into the spin basis defined by the applied magnetic field, we have
\begin{equation}
	\tilde{H} = \big( U \oplus U \big) H \big( U^{\dagger} \oplus U^{\dagger} \big) .
\end{equation} 
We can decompose the Hamiltonian into the form $\tilde{H} = \tilde{H}_{0} + \tilde{V}$, where $\tilde{H}_{0}$ includes the bare Zeeman terms, residual exchange splitting $J_{11}$, valley splitting, and direct SOC coupling between the singlet $S$ and unpolarized triplet state $\tilde{T}_{0}$. The term $\tilde{V}$ encodes all other contributions from SOC.
The Hamiltonian $\tilde{H}_{0}$ has block form, where $h_{a}$ is a $2\times2$ matrix operating on the subspace $\mathcal{H}_{a}$ spanned by $\lbrace S, \tilde{T}_{0} \rbrace$ and $h_{b}$ is a diagonal $6\times6$ matrix operating on the subspace $\mathcal{H}_{b}$ spanned by all other states $\lbrace \tilde{T}_{-}, \tilde{T}_{+}, S^{*}, \tilde{T}_{0}^{*}, \tilde{T}_{+}^{*}, \tilde{T}_{-}^{*}  \rbrace$, 
\begin{equation}
\tilde{H}_{0} = \left(
\begin{array}{cc}
h_{a} & 0 \\
0 & h_{b}
\end{array}
\right),
\end{equation}
where 
\begin{eqnarray}
h_{a} = \left(
\begin{array}{cc}
0 & B \sin^{2}(\theta) \big( \tilde{\alpha}_{R} - \tilde{\alpha}_{L} - (\tilde{\beta}_{R} - \tilde{\beta}_{L}) \sin(2 \varphi)\big) \\
\cdot & J_{11}
\end{array}
\right)
\end{eqnarray}
and
\begin{equation}
h_{b} = \mathrm{diag} \left[ J_{11} + g \mu_{B} B, J_{11} - g \mu_{B} B, \Delta_{\mathrm{vs},R}, \Delta_{\mathrm{vs},R} + J_{11}, \Delta_{\mathrm{vs},R} + J_{11} + g \mu_{B} B, \Delta_{\mathrm{vs},R} + J_{11} - g \mu_{B} B\right] ,
\end{equation}
where we're using the shorthand notation
\begin{eqnarray}
\tilde{\alpha}^{L} & \equiv & (1+\cos(\varphi_{v,\mathrm{L}})) \alpha^{L} \nonumber \\
\tilde{\alpha}^{R} & \equiv & (1+\cos(\varphi_{v,\mathrm{R}})) \alpha^{R} \nonumber \\
\tilde{\beta}^{L} & \equiv & (1+\cos(\varphi_{v,\mathrm{L}})) \beta^{L} \nonumber \\
\tilde{\beta}^{R} & \equiv & (1+\cos(\varphi_{v,\mathrm{R}})) \beta^{R}.
\end{eqnarray}

Similarly, the perturbation $\tilde{V}$ is given in block form by
\begin{equation}
\tilde{V} = \left(
\begin{array}{cc}
0 & v_{ab} \\
v^{\dagger}_{ab} & v_{b} 
\end{array}
\right),
\end{equation}
where $v_{ab}$ is a $2\times6$ matrix encoding the SOC between the subspaces $\mathcal{H}_{a}$ and $\mathcal{H}_{b}$ and $v_{b}$ is a $6\times6$ matrix describing the SOC within the subspace $\mathcal{H}_{b}$.
To describe the effect of $\tilde{V}$ on the spectrum of the two-level subspace of interest, we perform a Schrieffer-Wolff transformation. This entails finding an anti-Hermitian operator $S$ such that the unitary transformation $H' = e^{S} \tilde{H} e^{-S}$ eliminates the effective coupling between our two-dimensional subspace of interest and all other states. Expanding this transformation in $S$, we have $H' = \tilde{H}_{0} + \tilde{V} + [S,\tilde{H}_{0}] + [S,\tilde{V}] + \frac{1}{2}[S, [S, \tilde{H}_{0}]] + \frac{1}{2}[S,[S,\tilde{V}]]$. Hence, to eliminate the off-diagonal block component $v_{ab}$ to lowest order in $S$, we require $[S,\tilde{H}_{0}] = -\tilde{V}$. First, we make the ansatz that $S$ only has nonzero components on the off-diagonal block so that
\begin{equation}
S = \left(
\begin{array}{cc}
0 & s_{ab} \\
s_{ba} & 0
\end{array}
\right),
\end{equation}
where $s_{ab}$ is a $2\times6$ matrix and $s_{ba} = -s_{ab}^{\dagger}$ due to $S$ being anti-Hermitian. 
Making use of the block structure of these Hamiltonian terms, we obtain the condition $h_{a} s_{ab} - s_{ab} h_{b} = v_{ab}$. Noting that $h_{b}$ is invertible for the range of parameters of interest here, we can rewrite this as a recurrence relation to obtain the form
 \begin{equation}
 s_{ab} = -\Big(\sum_{n=0}^{\infty} h_{a}^{n} v_{ab} h_{b}^{-n}\Big) h_{b}^{-1}.
\end{equation}
Retaining only the leading order in $h_{b}^{-1}$, we obtain $s_{ab} \approx -v_{ab}h_{b}^{-1}$. Using this to evaluate the desired $2\times2$ block of $H'$, we obtain
\begin{eqnarray*}
h_{a}' & = & h_{a} +\frac{1}{2} s_{ab} v_{ab}^{\dagger} +\frac{1}{2} v s_{ab}^{\dagger} \\
& \approx & h_{a} - v_{ab} h_{b}^{-1} v_{ab}^{\dagger}.
\end{eqnarray*}

In our case, we have
\small
\begin{equation*} 
v_{ab}^{\dagger} = \frac{B \sin(\theta)}{\sqrt{2}} \left(
\begin{array}{cc}
-\Delta \tilde{\alpha} \cos\theta - \big(\sin(2 \varphi) \cos\theta + i \cos(2 \varphi) \big) \Delta \tilde{\beta} & -\sigma \tilde{\alpha} \cos\theta + \big( \sin(2 \varphi) \cos\theta -i \cos(2 \varphi)\big) \sigma \tilde{\beta} \\

\Delta \tilde{\alpha} \cos\theta - \big(\sin(2 \varphi) \cos\theta + i \cos(2 \varphi) \big) \Delta \tilde{\beta} & -\sigma \tilde{\alpha} \cos\theta + \big(\sin(2 \varphi) \cos\theta + i \cos(2 \varphi)\big) \sigma \tilde{\beta}\\

0 & i \sqrt{2} \sin\theta \big( \alpha_{R} - \beta_{R} \sin(2 \varphi) \big) \sin(\varphi_{v,\mathrm{R}}) \\

i \sqrt{2} \sin\theta \big( \overline{\alpha}_{R} \! - \! \overline{\beta}_{R} \sin(2 \varphi) \big) & -\sqrt{2} \sin\theta \cos\theta \Big( i \overline{\alpha}_{R} + e^{-2 i \varphi} \overline{\beta}_{R} \Big) \\

\overline{\beta}_{R} \cos(2 \varphi) \! - \! i \cos\theta \big( \overline{\alpha}_{R} - \overline{\beta}_{R} \sin(2 \varphi) \big) & \Big[ \frac{e^{2 i \varphi} + e^{-2 i \varphi} \cos(2 \theta)}{2} \! + \! i \! \sin(2 \varphi) \! \cos\theta\Big] \overline{\beta}_{R} \! - \! i \overline{\alpha}_{R}\big(\! \sin^{2}\theta \! - \! \cos\theta\big) \\

\overline{\beta}_{R} \cos(2 \varphi) \! + \! i \cos\theta \big( \overline{\alpha}_{R} - \overline{\beta}_{R} \sin(2 \varphi) \big) & \Big[ \! - \! \frac{e^{2 i \varphi} + e^{-2 i \varphi} \cos(2 \theta)}{2} \! + \! i \! \sin(2 \varphi) \! \cos\theta\Big] \overline{\beta}_{R} \! +\! i \overline{\alpha}_{R}\big(\! \sin^{2}\theta \! - \! \cos\theta\big)

\end{array}
\right),
\end{equation*}
\normalsize
where $\Delta \tilde{\alpha} \equiv \tilde{\alpha}_{R} - \tilde{\alpha}_{L}$, $\Delta \tilde{\beta} \equiv \tilde{\beta}_{R} - \tilde{\beta}_{L}$ and $\overline{\beta}_{R} \equiv \beta_{R} \sin(\varphi_{v,\mathrm{R}})$, $\overline{\alpha}_{R} \equiv \alpha_{R} \sin(\varphi_{v,\mathrm{R}})$. Notice that $v_{ab}^{\dagger}$ vanishes completely when the magnetic field is applied normal to the interface ($\theta = 0$). Suppose the magnetic field is applied in-plane ($\theta = \pi/2$). For this family of cases, we have
\begin{equation}
v_{ab}^{\dagger} = \frac{B}{\sqrt{2}} \left(
\begin{array}{cc}
-i \cos(2 \varphi) \Delta \tilde{\beta} & -i \cos(2 \varphi) \sigma \tilde{\beta} \\
-i \cos(2 \varphi) \Delta \tilde{\beta} & i \cos(2 \varphi) \sigma \tilde{\beta} \\
0 & i \sqrt{2} \big( \overline{\alpha}_{R} - \overline{\beta_{R}} \sin(2 \varphi)\big) \\
i \sqrt{2} \big( \overline{\alpha}_{R} - \overline{\beta_{R}} \sin(2 \varphi)\big) & 0 \\
\overline{\beta}_{R} \cos(2 \varphi) & -i \overline{\alpha}_{R} \\
\overline{\beta}_{R} \cos(2 \varphi) & i \overline{\alpha}_{R}
\end{array}
\right).
\end{equation}
We now consider the two magnetic field orientations considered in this work, $\mathbf{B} \parallel [100]$ and $\mathbf{B} \parallel [110]$.
For the case $\mathbf{B} \parallel [100]$ ($\theta=\pi/2$, $\varphi = 0$) we have
\begin{equation}
v_{ab}^{\dagger} = \frac{B}{\sqrt{2}} \left(
\begin{array}{cc}
-i \Delta \tilde{\beta} & -i \sigma \tilde{\beta} \\
-i \Delta \tilde{\beta} & i \sigma \tilde{\beta} \\
0 & i \sqrt{2} \overline{\alpha}_{R} \\
i \sqrt{2} \overline{\alpha}_{R} & 0 \\
\overline{\beta}_{R} & -i \overline{\alpha}_{R} \\
\overline{\beta}_{R} & i \overline{\alpha}_{R}
\end{array}
\right).
\end{equation}
Similarly, for the case $\mathbf{B} \parallel [110]$ ($\theta=\pi/2$, $\varphi = \pi/4$) we have
\begin{equation}
v_{ab}^{\dagger} = \frac{B}{\sqrt{2}} \left(
\begin{array}{cc}
0 & 0 \\
0 & 0 \\
0 & i \sqrt{2} \big( \overline{\alpha}_{R} - \overline{\beta}_{R}\big) \\
i \sqrt{2} \big( \overline{\alpha}_{R} - \overline{\beta}_{R}\big) & 0 \\
0 & -i \overline{\alpha}_{R} \\
0 & i \overline{\alpha}_{R}
\end{array}
\right).
\end{equation}

We can now evaluate the ST rotation frequencies induced by SOC for these two cases.
For $\mathbf{B} \parallel [110]$ we have
\begin{eqnarray}
h_{a}' & = & \left(
\begin{array}{cc}
0 & B(\Delta \tilde{\alpha} - \Delta \tilde{\beta}) \\
B(\Delta \tilde{\alpha} - \Delta \tilde{\beta}) & J_{11} - \frac{B^{2} \overline{\alpha}_{R}^{2} \Delta_{\mathrm{vs},R}}{\Delta_{\mathrm{vs},R}^{2} - (g \mu_{B} B)^{2}}
\end{array}
\right) \\
& \approx & \left(
\begin{array}{cc}
0 & B(\Delta \tilde{\alpha} - \Delta \tilde{\beta}) \\
B(\Delta \tilde{\alpha} - \Delta \tilde{\beta}) & J_{11}
\end{array}
\right),
\end{eqnarray}
if $(B \overline{\alpha}_{R})^{2} / \Delta_{\mathrm{vs},R} \ll 1$. Hence,
\begin{equation}
\Delta_{[110]} \approx \sqrt{4 B^{2}(\Delta \tilde{\alpha} - \Delta \tilde{\beta})^{2} + J_{11}^{2}}.
\label{eq:Delta110}
\end{equation}
Similarly, for $\mathbf{B} \parallel [100]$ and assuming contributions only from Dresselhaus SOC, we find
\begin{equation}
\Delta_{[100]} \approx \sqrt{\Bigg( \frac{B^{2} \overline{\beta}_{R}^{2} \Delta_{\mathrm{vs},R}}{\Delta_{\mathrm{vs},R}^{2}-(g \mu_{B} B)^{2}} \Bigg)^{2} + \Bigg(\frac{2 B \big( \tilde{\beta}_{R}^{2} - \tilde{\beta}_{L}^{2}\big)}{g \mu_{B}}\Bigg)^{2}}.
\label{eq:Delta100}
\end{equation}
From these expressions, we can see the origin of the generally linear dependence of the rotation frequency on magnetic field magnitude for the [110] orientation and quadratic-like dependence for the [100] orientation. While the number of free model parameters here is too large to be fully constrained by available experimental data, we can obtain a reasonably good fit to the data by assuming vanishingly small Rashba SOC ($\vert \alpha_{L} \vert, \vert \alpha_{R} \vert \ll \vert \beta_{L} \vert, \vert \beta_{R} \vert$). This is consistent with \citet{jock2018}. 
To illustrate the underconstrained nature of the fit, in \figref{fig:Misfit}, we show how equivalently satisfactory fits may be obtained by allowing the valley phases $\varphi_{v,\mathrm{R}},\varphi_{v,\mathrm{L}}$ to differ while holding the underlying Dresselhaus couplings $\beta_{R},\beta_{L}$ to be equal or, alternatively, the converse. We note that the above \eqnref{eq:Delta100} reduces to the main text \MainTextEqnSecondorderterm{} for $\Delta_{\mathrm{vs},R} \approx g \mu_{B} B$ and similar Dresselhaus strengths $|\tilde\beta_L|\approx |\tilde\beta_R|$. We emphasize that while the model doesn't uniquely identify some microscopic parameters like the valley phases, the conclusion is that both intravalley and intervalley processes are necessary to explain the data of \RefFigFreqDep{}.

\begin{supfigure}[htbp]
		\centering
		\includegraphics[width=8cm]{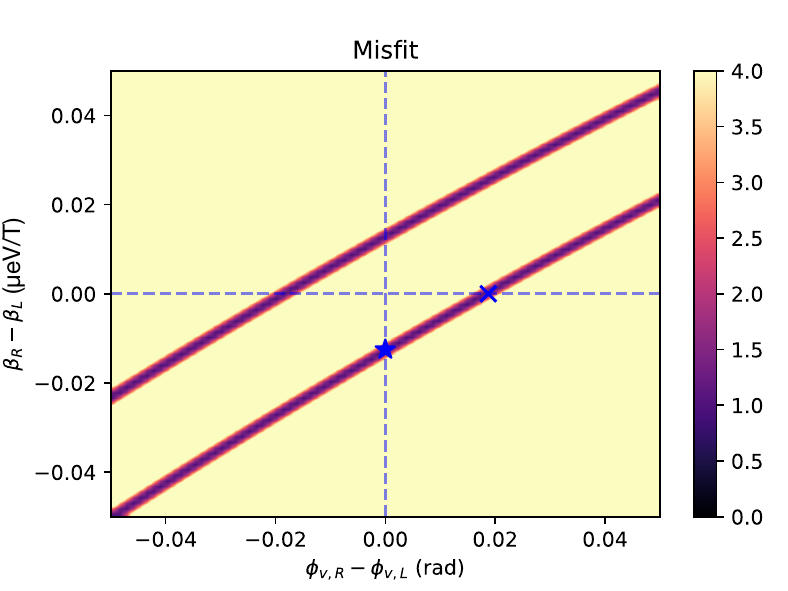}
		\caption{Misfit ($l^{2}$ norm of difference) between model and measured values for the $S{-}T_{0}$ rotation magnetic field dependence. Here, we are setting $\beta_{R}=0.7 \ \mathrm{\mu eV/T}$, $\varphi_{v,\mathrm{R}}=\pi/2$, and $\alpha_{L}=\alpha_{R}=0$. The points labeled by $\times$ ($\star$) indicate parameter choices where the misfit is the same but the valley phases (Dresselhaus couplings) are allowed to differ while the Dresselhaus couplings (valley phases) are set to be the same, respectively. For visual clarity, misfits larger than the maximum plotted have been clipped.}
		\label{fig:Misfit}
\end{supfigure}

\subsection{Model parameters known directly from experiment}

The \tabref{tab:modelparams} lists the parameters used to fit the experimental data.

{\centering
\begin{suptable}[htbp]
\begin{tabular}[c]{| c || c | c |}
\hline
 & $\mathbf{B} \parallel [100]$ & $\mathbf{B} \parallel [110]$ \\
\hline
Triplet-singlet splitting in $(2,0)$, $J_{20}$ ($\mathrm{\mu eV}$)& 243 & 255 \\
\hline
Right dot valley splitting, $\Delta_{\mathrm{vs},R}$ ($\mathrm{\mu eV}$) & 185 & Not characterized \\
\hline
BC interdot lever arm ($\mathrm{\mu eV/mV}$) & 90 & 90 \\
\hline
BC voltage at zero detuning ($\mathrm{V}$) & 2.9321 & 2.9533 \\
\hline
Singlet tunnel coupling, $t_{S}$ ($\mathrm{\mu eV}$) & 10 & 23\\
\hline
Triplet tunnel coupling, $t_{T}$ ($\mathrm{\mu eV}$) & 10 & 25 \\
\hline
\end{tabular}
\caption{Known model parameters, estimated from spin funnel and magnetospectroscopy measurements. The valley splitting in the right dot was not independently determined for the [110] magnetic field orientation, but has been assumed in our parameter fitting to be similar to the [100] case. Tunnel couplings here are full gap.}
\label{tab:modelparams}
\end{suptable}
\par}


\twocolumngrid

\end{document}